\documentclass[
aps,prd,   
               preprintnumbers,numbers,sort&compress,
               nofootinbib,
                            showpacs,
               colorlinks,
               linkcolor=blue,
               citecolor=blue]{revtex4-1} 

\usepackage{graphicx}
\usepackage{bm}
\usepackage{hyperref}

\usepackage{amsmath}
\usepackage[caption=false]{subfig}

 \def\ra{\rangle}
\def\la{\langle}

\newcommand{\be}{\begin{eqnarray}}
\newcommand{\ee}{\end{eqnarray}}
\newcommand{\beq}{\begin{equation}}
\newcommand{\eeq}{\end{equation}}
\newcommand{\exclude}[1]{}

\newcommand{\jcap}{JCAP}
\newcommand{\mnras}{Mon.\ Not.\ R.\ Astron.\ Soc.}
\newcommand{\aap}{A \& A }
  \exclude{

}
\graphicspath{{./Figures/}}

\begin{document}
       \title{The Mysterious  Bursts   observed by   Telescope Array  and    Axion Quark Nuggets }
       \author{Ariel  Zhitnitsky}
       \email{arz@phas.ubc.ca}
       \affiliation{Department of Physics and Astronomy, University of British Columbia, Vancouver, V6T 1Z1, BC, Canada}
   
      \begin{abstract}
       Telescope Array (TA) experiment has recorded  \cite{Abbasi:2017rvx,Okuda_2019}  several short time bursts of air shower like events. 
       These bursts are very distinct from conventional cosmic ray single showers, and  are found to be strongly correlated with lightnings. 
       We propose that these bursts  represent  the direct manifestation of the
    dark matter annihilation events within  the so-called axion quark nugget (AQN) 
    model, which was originally invented for completely different purpose to explain the observed  similarity between the dark   and the visible components  in the Universe, i.e. $\Omega_{\rm DM}\sim \Omega_{\rm visible}$ without any fitting parameters.  We  support this proposal by demonstrating that  the  observations  \cite{Abbasi:2017rvx,Okuda_2019}, including the  frequency of appearance, temporal and spatial distributions, intensity,  and  other  related  observables are  nicely match the emission features  of the AQNs propagating in the atmosphere under thunderstorm. We propose to test these ideas by reanalyzing the existing  data by increasing  the cutoff time scale $\Delta t=1$ ms for the bursts. We also suggest to test this proposal  by analyzing  the   correlations with proper infrasound and seismic instruments. We also suggest to search  for the radio signal with frequency $\nu\in (3-300)$ MHz which must be synchronized with  bursts.  
 \end{abstract}
     
       \maketitle

\section{Introduction}\label{sec:introduction}
In this work we discuss two naively unrelated stories.    
The  first one is  the study  of  a specific dark matter (DM) model, the  so-called axion quark nugget (AQN) model  \cite{Zhitnitsky:2002qa}, see a brief overview of this model below. The second one deals
with  the recent puzzling  observations \cite{Abbasi:2017rvx,Okuda_2019} by the Telescope Array (TA) experiment of the several short time bursts of air shower like events 
 as recorded by Surface particle Detector (TASD). 
The    bursts are defined as the events when at least three air shower triggers were recorded within 1 ms. Ten bursts have been recorded during   five years of the observations  (between May 11, 2008 and May 4,  2013). These bursts are very distinct from single showers    resulting from conventional  ultra high energy cosmic  ray (HECR)   events. The unusual  features are listed below \cite{Abbasi:2017rvx,Okuda_2019}:

{\bf 1}. Bursts start at much lower altitude than that of   conventional HECR showers.  All reconstructed air shower fronts for the burst events are much more curved than usual CR air showers;

{\bf 2}. Bursts events do not have sharp edges in waveforms as conventional HECR normally do;

{\bf 3}. The events are temporally clustered within 1 ms, which would be a highly unlikely occurrence for three consecutive conventional HECR  hits  in the same area within a radius of approximately 1km. The authors of    \cite{Abbasi:2017rvx} estimate the expectation of chance coincidence is less than $10^{-4}$ for five years of observations.

{\bf 4}. If one tries to fit the observed bursts with conventional code, the energy for HECR events should be in $10^{13}$ eV energy range, while the observed bursts correspond to $(10^{18}-10^{19})$ eV energy range as estimated by signal amplitude and distribution. Therefore, the estimated energy from individual events within the bursts is five to six orders of magnitude higher than the energy estimated by event rate;

{\bf 5}. Most of the observed bursts are ``synchronized" or ``related" with the lightning events, see precise definitions  below. Furthermore, all ten recorded bursts occur under thunderstorm;

{\bf 6}.  All bursts occur at the same time of lightning or earlier than lightning. Therefore, the bursts are associated with very initial moment of the lightning flashes, and  cannot be an  outcome nor consequence of the  lightning flashes as it would be detected at the later stages of   lightning flashes, not initial moment as observed;

{\bf 7}. The total 10 burst events have  been observed  during 5 years of observations. These bursts are not likely due to chance coincidence between single shower events.

The distinct features as listed above suggest that  the bursts are entitled to be considered as very puzzling rare events as they cannot be reconciled  with conventional CR, and we coin them as ``mysterious bursts".

  In this work we present the arguments suggesting that 
 these two  naively unrelated 
 things (AQN dark matter  \cite{Zhitnitsky:2002qa} and   the bursts as recorded by TASD \cite{Abbasi:2017rvx,Okuda_2019}) may   in fact be  intimately linked. In other words, we shall argue that ``mysterious bursts"  reported in  \cite{Abbasi:2017rvx,Okuda_2019}  could be a manifestation of the dark matter AQNs   traveling in the atmosphere during the thunderstorms. Our arguments are based on analysis  of the event rate, the energetics, the flux estimates, the time durations, the spatial distribution and other observables as recorded by  \cite{Abbasi:2017rvx,Okuda_2019}.   Therefore, we identify 
 the    ``mysterious bursts"  with  the AQNs hitting the Earth's    atmosphere under  the thunderstorms. We will show that all unusual features as listed in items {\bf 1-7 } above can be naturally explained  within the AQN framework. 
 
Now we are highlighting  the basic features of the AQN model which represents the first part of our story.
 The   AQN dark matter  model  \cite{Zhitnitsky:2002qa} was  invented long ago with a single motivation 
 to naturally explain the observed  similarity between the dark matter and the visible densities in the Universe, i.e. $\Omega_{\rm DM}\sim \Omega_{\rm visible}$ without any fitting parameters. 
 The AQN construction in many respects is 
similar to the original quark-nugget model suggested  by Witten \cite{Witten:1984rs}, see  \citep{Madsen:1998uh} for a review. This type of DM  is ``cosmologically dark'' not because of the weakness of the AQN interactions, but due to their small cross-section-to-mass ratio, which scales down many observable consequences of an otherwise strongly-interacting DM candidate. 

There are two additional elements in the  AQN model compared to the original models \cite{Witten:1984rs,Farhi:1984qu,DeRujula:1984axn,Madsen:1998uh}. First new element is the presence of the  axion domain walls which   are copiously produced  during the  QCD  transition. This domain wall plays a dual role: first  it serves  as  an   additional stabilization factor for the nuggets,     which helps to alleviate a number of  problems with the original nugget construction  \cite{Witten:1984rs,Farhi:1984qu,DeRujula:1984axn,Madsen:1998uh}.  Secondly, the same axion field $\theta (x)$ generates the strong $\cal{CP}$ violation in the entire visible Universe. This is because the $\theta (x)$ axion field before the QCD epoch could be thought as   classical $\cal{CP}$ violating field correlated on the scale of the entire Universe.
The axion field starts to oscillate at the QCD transition by emitting the propagating axions. However, these oscillations remain coherent on the scale     of the entire Universe. Therefore, the $\cal{CP}$ violating phase remains coherent on the same enormous scale.

  Another feature of the  AQN model which plays an
absolutely crucial role for  the present work  is that nuggets can be made of {\it matter} as well as {\it antimatter} during the QCD transition. 
Precisely the coherence of the $\cal{CP}$ violating field on large scale mentioned above provides a preferential production of one species of nuggets made of {\it antimatter}   over another 
species made of {\it matter}. The preference    is determined by the initial sign of the $\theta$ field when the formation of the AQNs starts.  
The direct consequence of this feature along with coherent $\cal{CP}$ violation in entire Universe  is that  the DM density, $\Omega_{\rm DM}$, and the visible    density, $\Omega_{\rm visible}$, will automatically assume the  same order of magnitude densities  $\Omega_{\rm DM}\sim \Omega_{\rm visible}$ without any fine tuning, see next section \ref{AQN-flux} with more details. 

One should emphasize that AQNs are absolutely stable configurations on cosmological scales. Furthermore, the antimatter which is hidden  in form of the very dense nuggets is unavailable for annihilation unless the AQNs hit the stars or the planets. 
There are also very rare events of annihilation in the center of the galaxy, which, in fact, may explain some observed galactic excess emissions in different frequency bands.
 Precisely the AQNs made of antimatter are capable to release a significant amount of  energy when they enter the Earth's atmosphere    and annihilation processes start to occur between  antimatter hidden in form of the AQNs and the   atmospheric  material. The term ``annihilation" in this work implies the elementary annihilation event between the antiquark from the bulk of the AQNs and the quark from visible baryons from surrounding material when the direct collision between AQN and the visible baryons occur. The annihilation between the positrons from the AQNs and electrons from visible matter also occur. However, the corresponding lepton processes  generate much less energy, and can be ignored. 
 
 It is important  to  comment here  that the thunderclouds 
 play a crucial   role in our discussions  as they serve as the triggers which  greatly increase the particle  emission  rate from the AQNs. 
This is because  the thunderclouds are  characterized by large preexisting electric field which serves as a trigger and  accelerator of the liberated positrons. 
 Precisely these features  which occur under thunderclouds  explain why all the recorded mysterious bursts  are  observed exclusively in the presence of the thunderclouds. 
 The necessity to have thunderclouds in the area also explains why 
 the   ``mysterious bursts" are so rare events: the AQNs (which are much more frequent events by themselves, see  below) must enter the area  under the thunderclouds    to   accelerate and   intensify   the emission  of the positrons  which will be  eventually recorded by TASD. 
 
  We conclude this Introduction with the following  comment. The annihilation events which inevitably occur when AQN interact with environment lead to many observable effects due  to release of a large amount of energy. 
In particular, the  corresponding annihilation events when AQN enters the Earth atmosphere lead to the release of energy in the form of the weakly coupled axions and  neutrinos as well as X and $\gamma$ rays, electrons, positrons and other particles.   It is hard to observe axions and neutrinos due to their feeble interactions, though the corresponding computations have been carried out recently, see  \cite{Lawson:2019cvy,Liang:2019lya} 
with the computations on the axion production and \cite{Zhitnitsky:2019tbh} with estimations of the AQN-induced neutrino flux. At the same time,  the X and $\gamma$ rays   emitted by AQNs are absorbed over distances $\sim 10$\,m or so in the atmosphere, and therefore cannot be easily recovered for analysis. The characteristic lifetime of free electrons is also very short and about $10^{-7}$s.  The liberated positrons, on the  other hand, can get accelerated under thundercloud and  propagate several kilometres in atmosphere at the sea level (and even much more at higher altitudes). 

We propose here  that  the AQN-induced positrons  are the source of the unusual  burst events observed by TASD    \cite{Abbasi:2017rvx,Okuda_2019}. 
We should emphasize that the AQN model was not designed nor invented to  explain the  ``mysterious bursts". Rather, the AQN model
was constructed  for completely different purposes, to explain the observed relation: $\Omega_{\rm DM}\sim \Omega_{\rm visible}$ without any fine tunings.  
As a consequence of the construction (manifested in a large amount of antimatter hidden in form of the AQNs)   this model also predicts a large number of positrons which will be liberated during the AQN traversing the atmosphere under the thunderstorm. These energetic positrons can mimic the CR air-shower, and we identify the  unusual TA burst   characterized by    items {\bf 1-7} listed above as a {\it cluster} of the events
generated by a single AQN.

Our presentation is organized as follows. In next section \ref{AQN-flux} we overview the basic ideas of the AQN model 
paying special attention to the specific topics relevant for the present studies. In section     \ref{proposal} we formulate the basic ideas of the proposal and make a number of  estimates including the  frequency of appearance, the emergence of clusters identified with bursts, estimate their intensity, etc. In section   \ref{confronts} we confront our proposal with observations \cite{Abbasi:2017rvx,Okuda_2019} by explaining how  the  unusual features listed  in items {\bf 1-7}
 naturally emerge in this AQN framework. Finally, in section \ref{conclusion} we formulate our basic findings and suggest possible tests  how this proposal  can be confirmed or refute by future studies.

\section{The AQN model.}\label{AQN-flux}
This section represents a relatively short overview of the AQN framework
where  we briefly mention  the basic ideas 
of the AQN model relevant for the present studies, especially in subsection \ref{TQ}. 
\subsection{The basics. Overview}
 The original motivation for this model can be explained in two lines as follows. 
It is commonly  assumed that the Universe 
began in a symmetric state with zero global baryonic charge 
and later (through some baryon number violating process, non- equilibrium dynamics, and $\cal{CP}$ violation effects, realizing three  famous   Sakharov's criteria) 
evolved into a state with a net positive baryon number.

 We advocate a model in which 
``baryogenesis'' is actually a charge segregation  (rather than charge generation) process 
in which the global baryon number of the universe remains 
zero at all times.  This scenario should be considered as  an 
alternative   path which is  qualitatively  distinct  from conventional baryogenesis.  
We    refer  to original works  \cite{Liang:2016tqc,Ge:2017ttc,Ge:2017idw,Ge:2019voa} devoted to the specific questions  related to the nugget's formation, generation of the baryon asymmetry, and 
survival   pattern of the nuggets during the evolution in  early Universe with its unfriendly environment. 

The result of this charge segregation   process is two populations of AQN carrying positive and 
negative baryon number. In other words,  the AQN may be formed of either {\it matter or antimatter}. 
However, due to the global  $\cal CP$ violating processes associated with $\theta_0\neq 0$ during 
the early formation stage,  the number of nuggets and antinuggets 
  will be different\footnote{The strong ${\cal CP}$ violation is related to the fundamental initial parameter $\theta_0$.
  This source of ${\cal CP}$ violation  is no longer 
available at the present time due to  the axion and its dynamics in early Universe. 
One should mention that the axion  remains the most compelling resolution of the strong ${\cal CP}$ problem, see original papers 
 on the axion \cite{axion1,axion2,axion3,KSVZ1,KSVZ2,DFSZ1,DFSZ2}, and   recent reviews \cite{vanBibber:2006rb, Asztalos:2006kz,Sikivie:2008,Raffelt:2006cw,Sikivie:2009fv,Rosenberg:2015kxa,Marsh:2015xka,Graham:2015ouw, Irastorza:2018dyq}.}.
 This disparity  is always an order of one effect   irrespectively to the 
parameters of the theory, the axion mass $m_a$ or the initial misalignment angle $\theta_0$.
 In this model the AQNs  represent 
the dark matter in the form of dense nuggets of  quarks (or antiquarks) and gluons in colour superconducting (CS) phase.

Furthermore, this disparity implies that  the total number of  visible antibaryons  will be less than the number of baryons in   early universe plasma, which is determined by the sign of the initial sign of $\theta_0\neq 0$.
These anti-baryons will be soon
annihilated away leaving only the visible baryons whose antimatter 
counterparts are bound in the excess of antiquark nuggets and are thus 
unavailable to annihilate.  As we previously mentioned, all asymmetries    are of order of 
one effects. This  is precisely  the reason why the resulting visible and dark matter 
densities must be the same order of magnitude as we already mentioned:
\be
\label{Omega}
 \Omega_{\rm DM}\sim \Omega_{\rm visible}
\ee
as they are both proportional to the same fundamental $\Lambda_{\rm QCD} $ scale,  
and they both are originated at the same  QCD epoch.
  If these processes 
are not fundamentally related the two components 
$\Omega_{\rm DM}$ and $\Omega_{\rm visible}$  could easily 
exist at vastly different scales.

 Another fundamental ratio (along with 
$\Omega_{\rm DM} \sim  \Omega_{\rm visible}$  discussed above)
is the baryon to entropy ratio at present time
\be
\label{eta1}
\eta\equiv\frac{n_B-n_{\bar{B}}}{n_{\gamma}}\simeq \frac{n_B}{n_{\gamma}}\sim 10^{-10}.
\ee
In our proposal (in contrast with conventional baryogenesis frameworks) this ratio 
is determined by the formation temperature $T_{\rm form}\simeq 41 $~MeV  at which the nuggets and 
antinuggets compete their formation, when all anti baryons get annihilated and only the baryons remain in the system. 
The $T_{\rm form}$ is very hard to compute theoretically as even the phase diagram for CS phase is not well known.  This temperature  of cosmic plasma   is  known from  the observed  ratio (\ref{eta1}).  However, we  note that $T_{\rm form}\approx \Lambda_{\rm QCD}$   assumes a typical QCD value, as it should as there are no any small parameters in QCD. 
We   refer to the original paper \cite{Zhitnitsky:2006vt}  with simple explanations  why this model  satisfies all the conventional for DM requirements such as decoupling from CMB, retaining  the standard picture of    structure formation and BBN  without any noticeable modifications, etc.  

\exclude{
It is known  that the galactic spectrum 
contains several excesses of diffuse emission the origin of which is not well established,  and  remains to be debated.
 The best 
known example is  the strong galactic 511~keV line. If the nuggets have a baryon 
number in the $\langle B\rangle \sim 10^{25}$ range they could offer a 
potential explanation for several of 
these diffuse components. It is very nontrivial consistency check that the required $\langle B\rangle$ to explain these excesses of the galactic diffuse emission  belongs to the same mass range as reviewed below. 

As we mentioned above this type of DM  is ``cosmologically dark'' not as a result of  weakness of the interaction, but as a result of small    cross-section-to-mass ratio, which scales down many observable consequences of an otherwise strongly-interacting AQNs. 
}
As we already mentioned the strongly interacting AQNs are dark due to the very small    cross-section-to-mass ratio. The observable effects do occur when DM and visible matter densities are sufficiently large and rare events of annihilation occur.  
In particular, the AQN model may  explain  some excesses of diffuse emission from the galactic center the origin of which  remains to be debated,   see the original works \cite{Oaknin:2004mn, Zhitnitsky:2006tu,Forbes:2006ba, Lawson:2007kp,Forbes:2008uf,Forbes:2009wg} with explicit  computations 
 of the galactic radiation  excesses  for varies frequencies, including excesses of the diffuse  X- and   $\gamma$- rays.  
In all these cases photon emission originates 
from the outer layer of the nuggets known as the electrosphere, and all intensities in different frequency bands are expressed in terms of a single parameter $\langle B\rangle $ representing the average baryon charge of the nuggets.  At present time this parameter can be inferred from observations. In next section we review a number of different observations which constraint this single fundamental parameter of the model. 

The  AQNs may also offer a resolution to some  seemingly unrelated puzzles such as   the ``Solar Corona Mystery"  \cite{Zhitnitsky:2017rop,Raza:2018gpb} when the   so-called ``nanoflares" conjectured by Parker long ago \cite{Parker} are   identified with the  annihilation events in the AQN framework.
    The luminosity  of the Extreme UV (EUV) radiation from corona due to these annihilation events is unambiguously determined by the DM density. It is very nontrivial consistency check that  the computed  luminosity from the corona nicely matches with observed EUV radiation.     The same   events  of annihilation are  also   manifested themselves as 
     the radio impulsive events \cite{Ge:2020xvf} in quiet solar corona. The computed  intensity and   spectral features    are consistent with recently recorded radio  events   in quiet solar corona  by Murchison Widefield Array Observatory \cite{Mondal-2020}.

  The  AQNs may also offer a resolution of   the     ``Primordial Lithium Puzzle" \cite{Flambaum:2018ohm} and  the longstanding puzzle with the DAMA/LIBRA  observation  of the annual modulation at $9.5\sigma$ confidence level  \cite{Zhitnitsky:2019tbh}. Furthermore, it may resolve the observed  (by XMM-Newton at  $ 11\sigma$ confidence level  \citep{Fraser:2014wja})  puzzling   seasonal variation   of the X-ray background in the near-Earth environment in the 2-6 keV energy range as suggested in  \cite{Ge:2020cho}. 
   The AQN annihilation events in the Earth's atmosphere could produce  infrasound and seismic acoustic waves     as discussed in   \cite{Budker:2020mqk}. 

   As we mentioned above  the  single fundamental parameter which essentially determines all the intensities for all the effects mentioned above is the average baryon charge $\la B \ra$ of the AQNs. There are a number of constraints on this parameter which are reviewed below. 
One should also mention that the AQN masses related to their baryon charge by $M_N\simeq m_p |B|$, where $m_p$ is the proton mass.  The AQNs are macroscopically large nuclear density objects.   For the present work we adopt a typical nuclear density of order $10^{40}\,{\rm cm^{-3}}$ such that, for example,  a nugget with $|B|\simeq 10^{25}$, which is a typical AQN's baryon charge as discussed in next section \ref{size},  has a   radius $R\simeq 2.2\cdot 10^{-5}{\rm cm}$ and mass of order 10 g. 

It should be contrasted with conventional meteors when an object  with a mass 10 g. would have a typical size of order 1 cm occupying the volume which would be  15 orders of magnitude larger than the AQN's volume. This is of course is due to the  fact that AQNs have nuclear density which is  15 orders of magnitude higher than   the density of a normal matter.   One can view an AQN as a very small neutron star (NS) with its nuclear density. The difference is that  the  NS is squeezed by the gravity, while the AQN is squeezed by the axion domain wall pressure. 
 \subsection{Size distribution. Frequency of appearance.}\label{size}

We now overview the observational constraints on such kind of dense objects which play a key role in our analysis
in identification of the   ``mysterious bursts"  recorded by  \cite{Abbasi:2017rvx,Okuda_2019}  
with the AQN annihilation events in atmosphere under the thunderstorm. 
 
The strongest direct detection limit\footnote{Non-detection of etching tracks in ancient mica gives another indirect constraint on the flux of   dark matter nuggets with mass $M< 55$g   \cite{Jacobs:2014yca}. This constraint is based on assumption that all nuggets have the same mass, which is not the case  as we discuss below.
The nuggets with small masses represent a tiny portion of all nuggets in this model.} is  set by the IceCube Observatory's,  see Appendix A in \cite{Lawson:2019cvy}:
\be
\label{direct}
\la B \ra > 3\cdot 10^{24} ~~~[{\rm direct ~ (non)detection ~constraint]}.
\ee
The upper limit on such kind of objects is a matter of debates as there are two computations leading to different results. We overview both results:  \cite{Herrin:2005kb} and  \cite{Cyncynates:2016rij}. The authors of   \cite{Herrin:2005kb}   use the Apollo data to constrain the abundance of  quark nuggets  in the region of 10\,kg to one ton.  The authors of  \cite{Herrin:2005kb}   argued that the contribution of such heavy nuggets  must be at least an order of magnitude less in order to  saturate the dark matter in the solar neighbourhood. Assuming that the AQNs do saturate the dark matter, the constraint  \cite{Herrin:2005kb} can be reinterpreted   that at least $90\%$ of the AQNs must have masses below 10\,kg. The authors of   \cite{Cyncynates:2016rij} argued that in fact,     there is no upper bound for such nuclear density nuggets\footnote{The difference with previous studies   is related to the accounting for  attenuation, geometric lensing and some other effects in \cite{Cyncynates:2016rij} which had been previously ignored in \cite{Herrin:2005kb}.}.
These results  can be approximately expressed  in terms of the baryon charge as follows:
   \be
\label{apollo}
\la B \ra &\lesssim &  10^{28} ~~~  [\rm Apollo~ analysis]   ,  \nonumber \\
\la B \ra & <  &  [\rm unconstrained] ~~~ [\rm Apollo~ analysis]      ,
\ee
depending on Apollo analysis of  \cite{Herrin:2005kb}  or   \cite{Cyncynates:2016rij} correspondingly.
For our purposes the difference between these two cases is not very important though as we will be using the models for the $B$ distribution which is 
motivated by studies of the so-called nanoflares as we discuss below.

Therefore, indirect observational constraints (\ref{direct}) and (\ref{apollo}) suggest that if the AQNs exist and saturate the dark matter density today, the dominant portion of them   must reside in the window: 
\be
\label{window}
3\cdot 10^{24}\lesssim \la B \ra \lesssim ~10^{28}   ~~~~~{ \rm  or} ~~~~~ [  \rm unconstrained   ] .
\ee

We emphasize that the AQN model within window in equation (\ref{window})  is consistent with all presently available cosmological, astrophysical, satellite and ground-based constraints.  Furthermore, it has been shown that these macroscopical objects can be formed, and the dominant portion of them will survive the dramatic events (such as BBN, galaxy and star formation etc) during the long evolution of the Universe.   
This model is  very rigid and predictive as there is no much flexibility nor freedom to modify any estimates mentioned above which have been carried out with one and the same set of parameters in drastically different environments  when  densities and  temperatures    span  many orders in magnitude.  

This very strong claim requires some clarifications. By saying that the model is very rigid and predictive we mean that the only fundamental  unknown parameter of the model
is $\langle B\rangle $ which should be fixed once and forever by the observations which are unambiguously identified with the AQN events. Such observations have not been established yet, and this  parameter remains a fundamental unknown parameter of the model. There are many other parameters which appear in the present work such as the internal temperature $T$ of the nugget,  parameter $\eta(T)$ which accounts for the number of liberated positrons when the AQN enters the thunderclouds, etc. These parameters are very hard (if possible) to compute from the first principles. However, this deficiency in computational power for complex system should not be interpreted  as a weak point of the AQN model itself as these parameters are not related to the model itself, but to complex behaviour of the AQNs when they enter a dense environment with large Mach number $M=v_{\rm AQN}/c_s\gg 1$ when the turbulence, shock waves and other complex phenomena occur\footnote{ The same comment obviously applies to the SM of particle physics which is undoubtedly  is very rigid and predictive model. However, a number of questions such as propagation of the meteoroids in atmosphere with supersonic speed, or explosion of a nuclear bomb which generates radiation and shock waves cannot be studied from first principles. Instead, the modelling is based on data analysis.}. In particular, in dilute environment such as galactic center, the relevant features of the AQNs for a given $B$ can be computed with high level of precision using such methods as Thomas-Fermi approximation, see \cite{Forbes:2008uf,Forbes:2009wg}.

  For our interpretation  of the   ``mysterious bursts"   \cite{Abbasi:2017rvx,Okuda_2019}   in terms of  the AQN annihilation events in atmosphere, one needs to know the size distribution and the frequency of appearance of AQNs with a given size.    The corresponding AQN flux is proportional to the dark matter number density $n_{\rm DM}v_{\rm DM}\sim (\rho_{\rm DM} v_{\rm DM})/\langle B\rangle$.  Therefore, 
 the corresponding  frequency of appearance  at which  the AQNs hit the Earth can be estimated as follows \cite{Lawson:2019cvy}: 
  \be
\label{eq:D Nflux 3}
\frac{\langle\dot{N}\rangle}{4\pi R_\oplus^2}
\simeq  \frac{5\cdot 10^{-2}}{\rm km^{2}~yr}
\left(\frac{\rho_{\rm DM}}{0.3{\rm \frac{GeV}{cm^3}}}\right)
 \left(\frac{\langle v_{\rm AQN} \rangle }{250~{\rm \frac{km}{s}}}\right) \left(\frac{10^{25}}{\langle B\rangle}\right),~~~~~
\ee
where we assumed the conventional galactic halo model with the local dark matter density being  $\rho_{\rm DM}\simeq 0.3\,{\rm  {GeV} {cm^{-3}}}$.  The number density of the AQNs is very tiny: $n_{\rm AQN}\simeq \left(\frac{\rho_{\rm DM}}{m_p\langle B\rangle} \right)$ as it is suppressed by factor $B^{-1}$. This is precisely the main reason  for  the nuggets to behave as the  cold dark matter particles. 

Averaging over all types of AQN trajectories  with different masses $ M_N\simeq m_p|B|$,  with different incident angles and different initial velocities and the size distribution can be obtained ( see \cite{Lawson:2019cvy}).  However, the results   do not dramatically  vary with corresponding  parameters, and   we shall use (\ref{eq:D Nflux 3}) in our estimates which follow. 
 The result (\ref{eq:D Nflux 3}) suggests that 
the AQNs hit the Earth's surface with a frequency  approximately  once a day   per   $(100 ~\rm km)^2$ area.    The rate is expressed in terms of the average value $\langle B\rangle$ as defined below by   (\ref{eq:f(B)}). This rate is suppressed for large sized  AQNs 
when $B$ is much larger than the mean value $\langle B\rangle$ and it is enhanced for $B\lesssim \langle B\rangle$
as we discuss below. 

  The corresponding size distribution (and corresponding  frequency of appearance)  is  defined as follows:
 Let $\rm d N/\rm d B$ be the number of AQNs    which carry the baryon charge [$B$, $B+dB$].
The mean value of  the baryon charge $\langle B\rangle $  which enters (\ref{eq:D Nflux 3}) is defined as follows:
\begin{equation}
\label{eq:f(B)}
\langle B\rangle 
=\int_{B_{\rm min}}^{B_{\rm max} }  dB~B f(B), 
~~~~~ f(B)\propto B^{-\alpha}, 
\end{equation}
where $f(B)$ is a  properly normalized distribution  and $\alpha\simeq (2-2.5)$  is the power-law index.
One should mention  that the parametrization  (\ref{eq:f(B)}) in terms of the distribution function $f$ 
was originally introduced  by solar physicists   to fit the observed extreme UV radiation assuming that the  corona heating is saturated   by the so-called nanoflares  (conjectured by Parker  \cite{Parker}
 many years ago to resolve the ``Solar Corona Mystery").   In the original construction function $f$    describes the    energy distribution $f(E_{\rm nano})$ for the nanoflares  \cite{Benz-2002,Pauluhn:2006ut,Bingert:2012se}.   This scaling was literally adopted  in \cite{Zhitnitsky:2017rop,Raza:2018gpb}, where it was proposed   that  the   nanoflares can be  identified with AQN-annihilation events in the solar corona and the energy of the nanoflare $E_{\rm nano}$ is related to the baryon charge $B$ of the AQN as follows: $E_{\rm nano}\simeq 2 m_pc^2 B \ \simeq (3\cdot 10^{-3} \ {\rm erg})   B$. As a result the nanoflare energy distribution $f(E_{\rm nano})$ and the AQN baryon charge distribution $f(B)$
is  the same (up to proper normalization) function $f(E_{\rm nano})\propto f(B)$ as  proposed  in \cite{Zhitnitsky:2017rop}. 
\exclude{
The highly nontrivial element of this identification is that  the required energy interval for the    nanoflares  being  in the range  $ E_{\rm nano}   \simeq     ( 10^{21}  -  10^{26})~{\rm erg.}$ as studied by   solar physicists   largely overlaps  with  allowed interval    for the AQN baryonic charge window (\ref{window}) derived from drastically different constraints. 
} 
By adopting this identification with nanoflare models  the maximum value for $B_{\rm max}$ was assumed\footnote{One should note that our results are not very sensitive to an exact  cutoff value $B_{\rm max}=10^{28}$ (corresponding to the nanoflare energy $E_{\rm nano}\simeq 3\cdot 10^{25}$ erg) because of  relatively large power index $\alpha$ strongly suppressing the contribution of the  large nuggets.} to be 
$B_{\rm max}=10^{28}$ which is consistent with  both studies performed in   \cite{Herrin:2005kb} and \cite{Cyncynates:2016rij}.

Therefore, we consider  a total of 6 different models for $f(B)$ corresponding to different models from \cite{Pauluhn:2006ut, Bingert:2012se}. In Table \ref{table:2.2 mean B} we show the mean baryon charge $\langle B \rangle$ for each of the 6 models. In all our recent studies  we  only investigated parameters that give $\langle B\rangle\gtrsim 10^{25}$ because present constraints require $\langle B\rangle > 3\cdot 10^{24}$ according to (\ref{direct}). This means excluding two models: that with $B_{\rm min} \sim 10^{23}$ and that with $\alpha=2. 5$ and $\alpha=2$. In all other models $\langle B \rangle$ varies between $(10^{25}-10^{26})$ and we use for $\langle B \rangle \sim 10^{25}$ for the estimates in this work.

\begin{table} [h] 
	\caption{Values of the mean baryon charge $\langle B\rangle$ for different parameters of the AQN mass-distribution function.} 
	\centering 
	\begin{tabular}{c|ccc} 
		\hline\hline
		$(B_{\rm min},\alpha)$ & 2.5                 & 2.0                 & (1.2, 2.5)          \\ \hline
		$10^{23}$                   & $2.99\times10^{23}$ & $1.15\times10^{24}$ & $ {4.25\times10^{25}}$ \\ 
		$3\times10^{24}$                   & $8.84\times10^{24}$ & $2.43\times10^{25}$ & $ {1.05\times10^{26}}$ \\ \hline
	\end{tabular}
	\label{table:2.2 mean B} 
\end{table}
The last column  in the table  with $\alpha=(1.2, 2.5) $ corresponds to   the nanoflare model considered by solar physics researchers when a combination of different exponents with $\alpha=1.2$ and $\alpha=2.5$ have been used  to describe separately  the  low energy and high energy   events to better fit the observations. 

The highly nontrivial element of this identification proposed in \cite{Zhitnitsky:2017rop,Raza:2018gpb} is that   the observed intensity of the extreme UV emission from the solar corona matches very nicely with  the total energy released as a result of   the AQN-annihilation events in the transition region.   One should emphasize that this ``numerical coincidence"  
 is a highly nontrivial self-consistency check of this proposal  connecting the conjectured solar nanoflares with AQNs, since the nanoflare properties  are  constrained by solar corona-heating models, while the intensity of the observed extreme UV due to the AQN annihilation events in the AQN framework  is mostly determined by the dark matter density $\rho_{\rm DM}\simeq 0.3\,{\rm  {GeV} {cm^{-3}}}$.
 
\subsection{ AQN's internal temperature $T$ and the  electric charge $Q$}\label{TQ}
  Another important element   relevant for our interpretation  of the   ``mysterious bursts"   \cite{Abbasi:2017rvx,Okuda_2019}   in terms of  the AQN annihilation events in atmosphere is the internal temperature $T$ of the nugget and its induced electric charge $eQ$. The corresponding parameters have been used 
  in our previous applications within AQN framework such as the  ``Primordial Lithium Puzzle" \cite{Flambaum:2018ohm},  the  solar corona heating puzzle \cite{Zhitnitsky:2017rop,Raza:2018gpb} and the seasonal variations observed by XMM-Newton   in x-ray frequency bands  \cite {Ge:2020cho}. In all the  previous cases the environment was drastically different from our present application of propagating the AQNs in the Earth's atmosphere. Nevertheless, the basic concept on structure of the nugget's core and its electrosphere, including the estimates of the  temperature $T$ and  electric charge $eQ$ remain the same.
  
  In context of the present work  the most relevant studies were  performed in   \cite{Budker:2020mqk} devoted to the acoustic signals generated by AQNs propagating in the Earth's atmosphere.   It has been speculated that some mysterious explosions when the infrasound and seismic acoustic waves have been recorded   can be identified with very rare large sized AQNs hitting the Earth. The estimations of the parameters $T$ and $Q$ from that analysis  \cite{Budker:2020mqk} can be directly applied to our present studies of the   ``mysterious bursts" recorded by TASD. The difference is that the paper     \cite{Budker:2020mqk} was  focused  on x-rays emission (with very short mean free path measured in meters at the sea level) which eventually generates the acoustic waves   propagating  for much longer distances. 
  
  In present studies we will be mostly interested in the AQN's emission of the positrons which can propagate few kilometers before they reach TASD to be recorded. However, the basic parameters  $T$ and $Q$ from   \cite{Budker:2020mqk}  relevant for  our present studies  remain basically the same. We highlight  the basic  ideas  on estimations of the temperature in Appendix \ref{sect:temperature} where some specific details relevant for the present work (such as the ionization features of the atmosphere  under thunderstorms at relatively high altitude $\sim 10$ km) are explicitly taken into account.  
 
 Another difference with    \cite{Budker:2020mqk} is that the main  focus in  the acoustic studies  was on  very rare and very intense events [which approximately occur once every ten years on $(100 \rm ~km)^2$ area]. These rare events are associated with very large nuggets with $B\simeq 10^{27}$ 
   which could generate the   infrasound  signal    being  sufficiently strong with overpressure on the level of $\delta p\simeq 0.3 $ Pa above the   detector's sensitivity. 
 Such  intense events are  very rare  ones according to (\ref{eq:f(B)}). In our present studies we are interested in typical and much more frequent  events with   $B\simeq 10^{25}$ when the  event rate is estimated by (\ref{eq:D Nflux 3}). The long-ranged    reveal    of the AQN interaction   in our present case will  be manifested     in form of the emitted positrons which can   propagate sufficiently long distance  as a result of  their acceleration by pre-exiting  electric field commonly present  during the thunderstorms.

 After an AQN hits  the dense region of the Earth's atmosphere, it acquires an internal temperature of the order $T\simeq (10-20)$ keV as a result of annihilation events in the nugget's core, see Appendix  \ref{sect:temperature}   for the details\footnote{One should note that only small portion of the antiquarks inside the nuggets get annihilated during entire journey of AQN crossing the Earth as there is no enough surrounding material which could directly hit the nugget. These nuggets will exit  the Earth's surface and continue their motion distancing from Earth with typical speed $v_{\rm AQN}\simeq 250 ~{\rm km/s}$.  The corresponding physics has been discussed in  \cite{Ge:2020cho} where it has been argued that the observed X-ray seasonal variation might be explained by these nuggets exiting the Earth.}.  Furthermore,  during its journey    the AQN's  speed $v_{\rm AQN}\simeq 250 ~{\rm km/s}$, which is   a typical DM  value,  greatly exceeds the speed of sound $c_s$ by several orders of magnitude such that Mach number $M=v_{\rm AQN}/c_s\gg 1$. The AQN temperature rise will cause the  excited positrons from 
 electrosphere   to expand well beyond the thin layer surrounding the nugget surface. Some of the positrons will  stay in close vicinity of the moving, negatively charged nugget core.  However, a finite portion of the positrons may leave the system  as a result of direct  elastic interaction  of the weakly coupled positrons with atmospheric molecules and due to the interaction with strong electric field which is always present in thunderclouds.  The number of weakly bound positrons $Q$ surrounding the nuggets at temperature $T$ can be estimated as follows:
\be
  \label{Q}
Q\simeq 4\pi R^2 \int^{\infty}_{0}  n(z, T)dz\sim \frac{4\pi R^2}{\sqrt{2\pi\alpha}}  \left(m_e T\right)   \left(\frac{T}{m_e}\right)^{1/4} , ~~
  \ee
where $n(z, T)$ is the positron number density in electrosphere,  which has been computed in the mean field approximation for the low temperatures  \cite{Forbes:2008uf,Forbes:2009wg}. The variable  $z$ in this formula describes the distance from the nugget's surface.
The electrosphere is normally very thin with typical width  much smaller than $R\sim 10^{-5}$cm, which justifies the description in terms of a simple flat  geometry  when the relevant  region of $|z|\ll R$. 
   
 In the equilibrium  with small  annihilation rate the positrons will occupy very thin layer  around the core's nugget as computed in \cite{Forbes:2008uf,Forbes:2009wg}. However, in atmosphere due to a large number of non-equilibrium processes such as generation of the shock wave (as a result  of  large Mach number $M$) and also due to  the  direct collisions with atmospheric molecules 
 the positron's cloud  is expected to expand well beyond the  thin layer  around the core's nugget. Some positrons will be kicked off and   leave the system. 
 
 How many positrons precisely will leave the system? This question is very hard  to answer in any quantitative way,     and we will assume that, to first order, that finite portion of them $\sim Q$ may leave the system as a result of shock wave and turbulence,  or as a result of  direct collisions with atmospheric molecules.  The remaining finite    portion of them $\sim Q$ will stay in the system and continues its motion with velocity $v_{\rm AQN}$  surrounding  the nugget's core\footnote{This case  should be contrasted with  our studies \cite {Ge:2020cho} of the seasonal variations observed by XMM-Newton due to the AQN's emission in X-ray frequency bands. The observations are performed   at large distances from Earth $r\sim  7 R_{\oplus}$ in empty space when weakly coupled positrons in electrosphere (\ref{Q}) cannot be kicked off as probability for  collisions is very tiny in empty environment. In this  case the dominant portion of the positrons $Q$ remains in the system all the time.}.  In that case,  the nugget core   acquires a negative  electric  charge $\sim -|e|Q$ such that  the nuggets get   partially  ionized.

 The distance $\rho$ at which the positrons remain attached to the nugget is given by the capture radius $R_{\rm cap}(T)$, determined by the Coulomb attraction:
\be
\label{capture}
\frac{\alpha Q(\rho)}{\rho} > \frac{m_e v^2}{2}\approx T ~~~{\rm for}~~~ \rho\lesssim  R_{\rm cap}(T),
\ee
where it is assumed that a finite portion of positrons  $\sim Q$ leave the system and finite portion of the positrons $\sim Q$ remain in the system, as explained above. The finite portion of the remaining positrons obviously cannot completely screen the initial charge $Q$. The  density of surrounding positively charged ions in thunderclouds $N_i\simeq (5-8) \cdot 10^4 {\rm cm^{-3}}$ as discussed in Appendix  \ref{sect:temperature}  is too small to play any role in possible screening effects.  

 The capture radius $R_{\rm cap}(T)$ is many orders of magnitude greater than nugget's size $R$ due to the long range Coulomb forces. The  binding energy represents the difference between Coulomb attraction and kinetic energy and must be obviously negative for the positrons to be tied  to the nugget.

We conclude this short overview  section with the following comment.  The rise  of the  temperature $T$ and consequently, the  electric charge $Q$ as discussed above is very generic feature   of the AQN framework when the nuggets  propagate in  atmosphere and annihilation processes occur in the nugget's core.  The observable effects will be drastically intensified  due to the AQN interaction with pre-existing electric field which is always present in thunderclouds.  The features  expressed by  formulae (\ref{Q})   and (\ref{capture}) will play a crucial role  in our analysis of the interaction of this positron's cloud with pre-existing electric field in thunderstorms.  

To be more precise, in  subsections \ref{sect:thunder} and \ref{sect:AQN-thunder} below we argue that the sufficiently strong electric  field  may liberate and consequently accelerate these positrons to the 10 MeV energy    range  such that these positrons  can easily propagate several  kilo-meters before they get  annihilated.   These energetic positrons  can be eventually  detected by TASD.

\section{ ``Mysterious bursts" as   the AQN annihilation  events}\label{proposal}
 
In this section we formulate the basic idea  of the proposal on how the ``mysterious bursts"  recorded by  \cite{Abbasi:2017rvx,Okuda_2019}  can be interpreted  in terms of  the AQN annihilation  events under thunderstorm.  After that we make the  estimates supporting every single element of this proposal. 

The idea goes as follows [we separated item (a) which is generic and not specific to the atmospheric conditions from  item (b) which applies exclusively to the  case of the thunderstorm]:

a) The AQN propagating  in atmosphere experiences a large number of annihilation events between antimatter quarks hidden in the AQN's core and surrounding material.
The annihilation processes raise the internal temperature of the AQN very much in the same way as discussed in  \cite{Budker:2020mqk}.  
In the  atmosphere the finite portion of the  positrons may be kicked off due to the elastic collisions with  atmospheric molecules, in which case the positrons may leave the system. However, we expect that the finite portion of the positrons  remain bound to the nugget at distances of order  $R_{\rm cap}$ determined by (\ref{capture}). 

b) If the AQN hits the area under thundercloud the weakly bound  positrons localized far away from the AQN's core at $R_{\rm cap}$   may be liberated by preexisting electric field $\rm {\cal{E}}\sim  kV/cm$ which is known to exist in thunderclouds.   As a result of the strong electric field the  positrons will be accelerated to the relativistic velocities and energies of 10  MeV on scales of 100 meters or so.
The mean free path for such energetic positrons is several km such that these positrons can reach TASD detectors.

 Our proposal is that precisely these energetic positrons generate   the ``mysterious bursts"  recorded by   \cite{Abbasi:2017rvx,Okuda_2019}. Below we present the arguments supporting this claim. We shall demonstrate that all highly unusual features   listed as items {\bf 1-7} in Introduction   find very natural explanation within  this proposal, which represents the topic of the separate section \ref{confronts}.

\subsection{Frequency of appearance} \label{sec:event_rate} 
Here we want to estimate the total number of events which TASD can record within the AQN proposal during 5 years of observations. 
One should emphasize from the very beginning that our estimates which follow are the order of magnitude estimates as there are many unknowns as we discuss in course of the text.
Furthermore, the observed 10 bursts during 5 years implies that statistical fluctuations could be essential.
Our crucial  arguments supporting the main  claim of this work 
are  based on   unique features  of the events (it will be  the topic of the separate section \ref{confronts})
 rather than on  estimation  of the frequency of appearance   which is the topic of the present section \ref{sec:event_rate}. Nevertheless, we think it is useful to have such an order of magnitude estimate which as we shall see below is consistent with the observed rate.

The starting point is the AQN flux (\ref{eq:D Nflux 3}) which determines the total number of bursts $N_{\rm bursts}$ during 5 years of collecting data:
  \be
\label{N} 
N_{\rm bursts}\simeq  \frac{5\cdot 10^{-2}}{\rm km^{2}~yr}  \cdot  ({\cal{A}}  )\cdot ({\cal{T}}  ) \cdot ({\cal{F}}   ),
\ee
where ${\cal{A}}$ is the effective  area, ${\cal{T}}$  is the data collection time, and finally  factor   ${\cal{F}}$ describes  the  fraction of time when the effective area ${\cal{A}}$ has been  under thunderclouds.
We start with the simplest parameter ${\cal{T}}\simeq 5$ years according to   \cite{Abbasi:2017rvx,Okuda_2019}. 
\exclude{Estimation of parameter ${\cal{F}}$ is based 
on compilation of the annual thunderstorm duration from 450 air weather system in USA as described  in 
 \cite{Gurevich:2004km}.   The corresponding estimates suggest that on average the thunderstorms last about $1\%$ of  time in each given area  \cite{Gurevich:2004km}. We adopt this estimate such that ${\cal{F}}\simeq 10^{-2}$. 
   }
   Estimation of parameter ${\cal{F}}$ is based on the number of detected lightning during  5 years (between May 2008 and April 2013) in the area which was 10073  \cite{Abbasi:2017rvx,Okuda_2019}.
   Assuming that a typical thunderstorm lasts  one hour and    produces $10^2$  lightnings     \cite{DWYER2014147} one can infer that the total time when  the TASD  area was under a  thunderstorm 
   is $10073\cdot  10^{-2} h\simeq  10^{2} h$ which represents approximately ${\cal{F}}\simeq 0.25\cdot 10^{-2}$.
    
The last  part   is the estimation of the effective area ${\cal{A}}$. One could naively use the area ${\cal{A}}\simeq 680 ~\rm km^2$ covered by the grid array of the SD detectors   \cite{Abbasi:2017rvx,Okuda_2019}. However, it would be a strong underestimation for the problem under consideration. The point is that 
the AQN trajectory very often has large inclination angle. Furthermore, according to   \cite{Abbasi:2017rvx,Okuda_2019} the criteria for ``related"  lightning is  that the time difference between burst and lightning is less than 200 ms. This time scale determines the maximal  size  $L_{\rm thunder}$  of a thunderstorm system,  under which  the AQN   emits the positrons  (which can eventually  reach the TASD detectors).  The ``related"  lightning  obviously occurs in a different location of the sky, but within the same thunderstorm system. This distance is    estimated as follows
\be
\label{L}
L_{\rm thunder}\simeq v_{\rm AQN}\cdot (200~ \rm ms)\simeq 250 ~{\rm \frac{km}{s}}\cdot 0.2 s\simeq 50~ km, ~~~
\ee
which is a size for  a typical  thunderstorm system, and  few times larger than the size of TASD. This estimate 
suggests that  effective area ${\cal{A}}$ within AQN framework can be represented  as ${\cal{A}}\simeq  \frac{1}{2} L_{\rm thunder}^2\simeq 1.3\cdot 10^3 \rm km^2$,
where factor $1/2$ accounts for the angular distribution of the AQNs with arbitrary inclination angles in computing the flux. 

Collecting all factors together we arrive to our final estimate for the frequency of appearance of the  burst-like events which TA collaboration could observe  during 5 years of collecting data:
  \be
\label{N1}
N_{\rm bursts}\simeq  \frac{5\cdot 10^{-2}}{\rm km^{2}~yr}  (1.3\cdot 10^3 \rm km^2  )  ( 5 yr )   ( 0.25\cdot 10^{-2}  )\simeq 1.~~~
\ee
  This estimate (\ref{N1}) should be compared with the observed 10 bursts recorded by  \cite{Abbasi:2017rvx,Okuda_2019}. Our estimate is one order of magnitude below the observed value. 
Nevertheless,  we consider a   similarity between these numbers as very encouraging sign   as all elements entering (\ref{N})  come from very different fields which  are  determined by very different physics (DM and thunderstorms).  We also consider  
  this order of magnitude estimate (\ref{N1})  as a highly nontrivial consistency check for the proposal as the basic numerical factor (\ref{eq:D Nflux 3}) entering (\ref{N}) depends on the AQN size distribution model and can easily deviate\footnote{In particular, the event rate will be two times   higher  than presented  in (\ref{N1}) for  specific power law $\alpha\simeq  2.5$   because the  smaller size nuggets with $B\lesssim \la B\ra$ are much more numerous  according to scaling  (\ref{eq:f(B)}) when the result is expressed in terms of $B$ rather than in terms of $\la B\ra$.} by factor 2 or so even for the fixed local dark matter density  
   $\rho_{\rm DM}\simeq 0.3\,{\rm  {GeV} {cm^{-3}}}$ which could also strongly deviate locally from its canonical value\footnote{There are numerous hints suggesting that $\rho_{\rm DM}$ locally in solar system could be much larger than it is normally assumed. We shall not elaborate on this matter in present work.}. The total number of the observed bursts (which is 10 in 5 years) is also could be a statistical fluctuation due to the low statistics. However, we shall not elaborate on these specific details in the present work as the  main claim of this work     as stated at the very beginning of this subsection is that our crucial  arguments  are  based on  analysis of the  unique features listed as $\bf 1-\bf 7$  of the events,  and which are hard to explain within conventional CR assumption. Some of the observables 
as discussed in Sect. \ref{confronts}   have pure geometrical and kinematical nature and represent essentially model independent predictions within AQN proposal, in contrast with the estimation (\ref{N}) of the normalization factor   which suffers  large numerical uncertainties.

  \subsection{Electric field in thunderclouds}\label{sect:thunder} 
In this subsection we overview the properties  of the pre-existing electric field which is always present in thunderclouds.  It  represents a short detour from our 
main topic. However the corresponding parameters play a key role in our arguments in following subsection \ref{sect:AQN-thunder} devoted to study of the AQNs under the thunderstorms.

We refer to review papers \cite{Gurevich_2001,DWYER2014147} devoted to study of the lightning where pre-existing electric field plays a crucial role in dynamics of the  lightning processes. While there is a consensus  on typical parameters of the electric field  which are important for the  lightning dynamics, the physics of the of the initial moment of lightning remains a matter of debate, and refs \cite{Gurevich_2001,DWYER2014147} represent  different views on this matter, see also
 references \cite{doi:10.1029/2011JA016494,doi:10.1029/2011JA017431,doi:10.1029/2011JA017487} where some specific elements of existing disagreement have been explicitly formulated  and debated. 

 For our purposes, however, the disputable elements do not play any role in our studies. Important  elements for the present work, which are not controversial,  are the temporal and spatial characteristics of  the electric field 
and their  values under thunderclouds. These parameters are     well established   and  are  not part of the debates, and we quote these parameters below.

 We start by quoting the so-called critical electric field ${\cal{E}}_c$ which must exist in thunderstorms for occurrence of runaway breakdown (RB in terminology  \cite{Gurevich_2001}) or  relativistic runaway electron avalanche (RREA in terminology \cite{DWYER2014147}):
\be
\label{E}
{\cal{E}}_c= (2.16-2.84)\rm  \frac{kV}{cm} \exp\left(-\frac{z}{h}\right), ~~~ h\simeq 8 ~km.
\ee
Such strong  (and even stronger) fields are routinely observed in atmosphere using e.g. balloon measurements. In our order of magnitude estimates which follow we use $\cal{E}\simeq  \rm \frac{kV}{cm}$ to simplify numerics. 
 Another important characteristic  is the avalanche scale $l_a$  and the corresponding time scale $\tau_a$ for the exponential growth, which are numerically estimated as  \cite{DWYER2014147,Gurevich_2001}:
\be
\label{l_a}
l_a\simeq (50-100)~ {\rm m} , ~~~~~ \tau_a\simeq \frac{l_a}{c} \sim (\rm fraction ~ of) ~\mu s.
\ee
The characteristic scale $l_a$ represents the minimum length scale when the exponential growth of  runaway  avalanche occurs. 
The spatial  scale $L_{\cal{E}}$ of a   electric field in thunderstorm must substantially exceed the scale $l_a$ for the exponential growth of the avalanche, i.e $L_{\cal{E}}\gg l_a$
as  argued  in  \cite{DWYER2014147,Gurevich_2001}. The scale $L_{\cal{E}}$ essentially determines  the allowed scale of the   inhomogeneity  and   non-uniformity of a fluctuating   electric field for  the exponential growth  to hold for sufficiently long time. However, for our purposes in our estimates which follow we shall use the relevant length scale for electric field $l_a\simeq 100$ m because we are not interested in studies of the runaway electrons    responsible for the lightning when    very large length scale $L_{\cal{E}}\gg l_a$ is required. In our case of the positrons the acceleration occurs on the avalanche scale $l_a$  after which the positrons leave the region of the electric field as $L_{\cal{E}}$ does not represent a straight line due to the fluctuations on the scale of $l_a$. It should be    contrasted with electrons in RB processes when the electrons represent an essential part of the dynamics of the developing the avalanche and adjust their trajectories. In this case the RB processes    require  much greater scale $L_{\cal{E}}$ to be operational.

\exclude{
The electric field obviously shows strong temporal  fluctuates   (in particular, as a result of  spatial  inhomogeneities $\sim L_E$ in the system) with time scale $\tau_E$.  One can estimate  the corresponding  parameter  $\tau_E$ by proper rescaling  $\sim ({L_E}/{l_a})$ in comparison with computed value for $\tau_a$:
\be
\label{tau_E}
\tau_\cal{E}sim\left( \frac{L_\cal{E}}{l_a} \right)\tau_a \sim ({\rm few})~\mu s .
\ee
}
One should emphasize that these small scale strong fluctuating electric fields should be distinguished from large scale slow fields with scale measured in km and lasting for minutes. Furthermore, these small scale    fields should be distinguished from fields which are induced  as a result  of the RB \cite{Gurevich_2001}
or (RREA   \cite{DWYER2014147}) processes when the runaway electrons represent an important contribution to the induced electric field\footnote{\label{trigger}It is known that the presence of the strong field (\ref{E}) is   required  for  RB processes to start. Another element which remains to be a matter of debates \cite{Gurevich_2001,DWYER2014147, doi:10.1029/2011JA016494,doi:10.1029/2011JA017431,doi:10.1029/2011JA017487} is the nature of the seeded particles which play the role of   trigger of the RB processes. Our original comment here is that  AQNs could  also play the role of a  trigger (in cases of rare   burst synchronized  events) along with conventional sources such as  the cosmic rays as advocated  in \cite{Gurevich_2001}.}. One should also note that (\ref{E}) gives the critical value for the   field  for the breakdown processes  to start. The generic field which occupy a dominant portion of the relevant area could be several times smaller in magnitude. The field $\cal{E}$ also strongly fluctuates in time and space and the corresponding regions could have an  arbitrary orientation   with respect to earth's surface, in particular in case of intracloud  lightnings.  The corresponding dynamics of the small scale fluctuating field is obviously very complex problem. It is not theoretically well understood, but can be studied using the observations and laboratory  experiments, see  \cite{DWYER2014147} for review.

To conclude this short detour on  lightning processes one should emphasize that it is not our goal to study this complicated physics. Rather, our goal is to overview some basic characteristics of the electric field such as ${\cal{E}}, l_a, \tau_{\cal{E}}$ which are known to be present in the atmosphere under the thunderstorm
because such phenomenon as lightning obviously exists in nature and the field (\ref{E}) plays the key role at the initial moment when RB processes   start. 
In the next   subsection  
we argue that these parameters:
 \be
 \label{parameters1}
 {\cal{E}}\simeq  {\rm \frac{kV}{cm}}, ~~~~  l_a \simeq 100 ~{\rm m},~~~~ \tau_{\cal{E}}\simeq 0.3 \mu s
 \ee
  nicely match the required parameters to explain  the observed ``mysterious bursts" observed by  \cite{Abbasi:2017rvx,Okuda_2019} which are interpreted in this work as the AQN annihilation events under thunderstorm.

\subsection{AQNs under   thunderstorm}\label{sect:AQN-thunder}

We are now prepared for the  analysis of the AQN weakly coupled positrons  (\ref{Q})   under influence of the pre-existing electric field characterized by parameters reviewed in   subsection \ref{sect:thunder}. 
As previously mentioned we expect that in the  atmosphere the finite portion of the  positrons may be kicked off due to the elastic collisions with  atmospheric molecules but  another (also finite) portion of the positrons  being also hit by heavy molecules still remain bound to the nugget at distances of order  $R_{\rm cap}$ determined by (\ref{capture}), which can be  numerically estimated as:
\be
\label{capture1}
R_{\rm cap}(T)\simeq \frac{\alpha Q}{T}\sim 2~{\rm cm} \left(\frac{T}{10 \rm ~keV}\right)^{1/4}.
\ee
At this distance the bound positrons are characterized by potential and kinetic energies (with opposite signs, of course) of order $T$. However, the absolute value of the binding energy $|E_{\rm bound}|\ll T$   as a result of strong cancellation between these two contributions.   Therefore, a sudden appearance  of strong external electric field (\ref{parameters1}) along the AQN's path will inject an additional energy $\Delta E$ estimated as
\be
\label{Delta_E}
\Delta E \simeq [e{\cal{E}}\cdot R_{\rm cap}]\sim 2   ~{\rm keV}\gtrsim E_{\rm bound} ~ {\rm at} ~ t=0.
\ee
 This additional energy injection  of order of several  keV could  liberate the weakly coupled positrons  from the nuggets. At this initial  moment $t=0$ the positrons will have kinetic energy of order $\sim \rm keV$. It is very important  that finite portion of the  weakly bound positrons (to be estimated below) will be liberated almost instantaneously at the same time  $t=0$ when the AQN enters the region with strong electric field.

\begin{figure}
    \centering
    \includegraphics[width=0.8\linewidth]{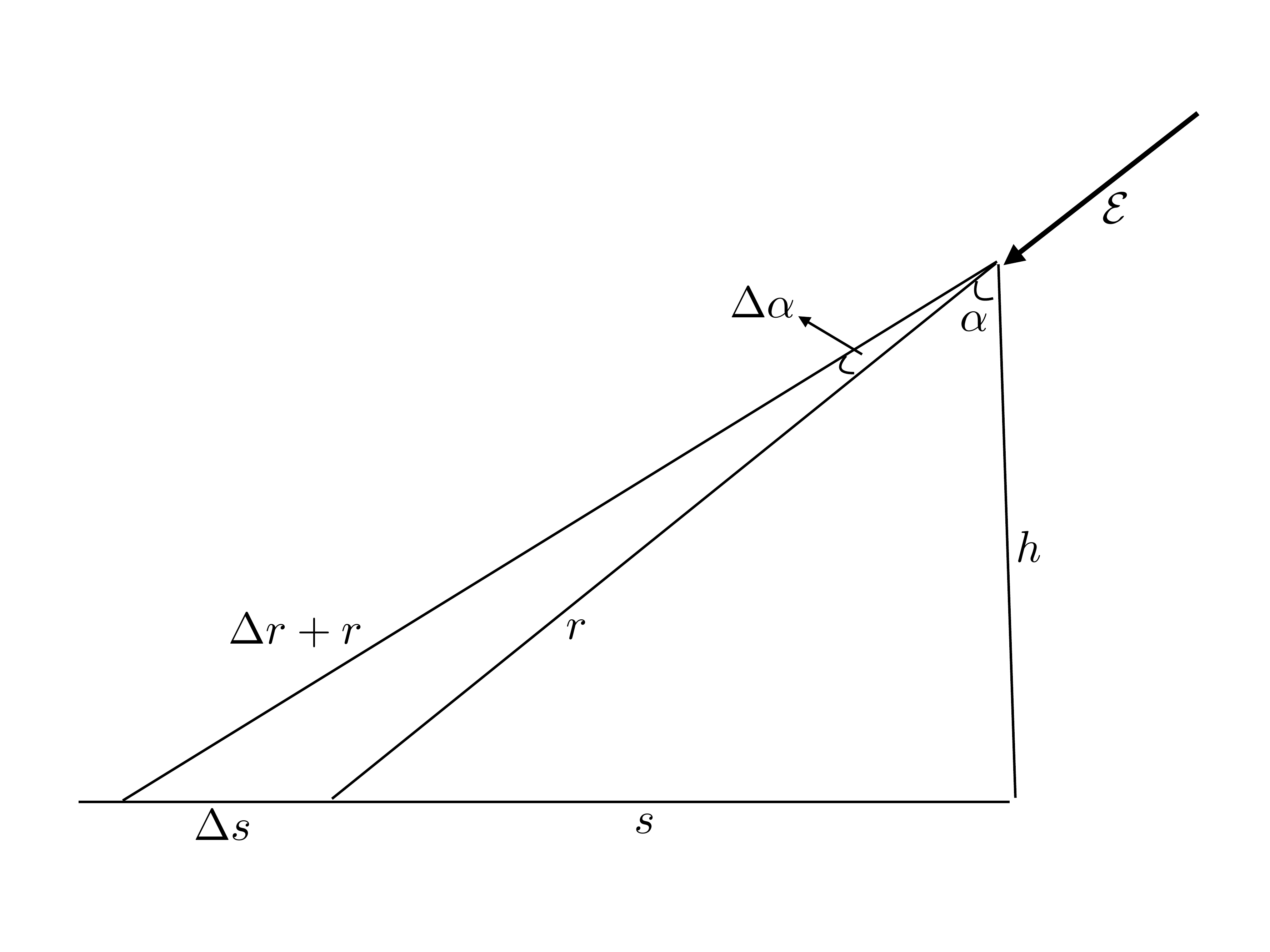}
    \caption{The positrons move along the cone between  path $``r"$ and $``r+\Delta r"$ with inclination angle $\alpha$ with respect to the vertical direction.  The angular spread  $\Delta \alpha\ll \alpha$ is assumed to be small. The spatial spread on the surface is determined  by $\Delta s$, while the additional travelling path is determined by $\Delta r$, see estimates in the text. The altitude is assumed to be within conventional range $ h\simeq (4-12)$ km. Instant  direction of the electric field $\vec{{\cal{E}}}$ at the moment of emission of the positrons   is also shown.}
    \label{geometry}
\end{figure}

 These liberated positrons find themselves in the background of strong electric field characterized by typical length 
 scale $l_a \simeq 100$ m according to (\ref{parameters1}).
 This pre-existing field will accelerate them to MeV energies on   scale (\ref{parameters1}). Indeed, 
 \be
\label{E_MeV}
  E_{\rm exit}\simeq [e{\cal{E}} \cdot l_a]\sim 10 \rm ~MeV 
\ee
after the positrons exit the region of the electric field. 
All suddenly released positrons will obviously   move   in the same direction   which is entirely determined by the direction of the electric field $\vec{{\cal{E}}}$ at the moment of exit.  Small 
angle $\Delta \alpha$ in the velocity 
  distribution   at the exit point is determined  by initial energy (\ref{Delta_E}) which is determined by  transverse  component perpendicular to electric field:    
$v_{\perp}\simeq \sqrt{2 \Delta E/m}$.   Therefore, after travelling the distance $r$  the   spatially spread  range $\Delta s $ is estimated as 
\be
\label{spread}
\Delta s \simeq  r \left(\frac{\Delta \alpha}{\cos \alpha}\right) \simeq \frac{1~{\rm km}}{\cos\alpha} \left(\frac{r}{10 \rm ~km}\right)\left(\frac{\Delta \alpha}{0.1}\right) , ~~~~~~ \Delta \alpha\simeq \left(\frac{v_{\perp}}{c}\right)\in (0- 0.1),   ~~~~ \Delta \alpha\ll \alpha, 
\ee
see Fig. \ref{geometry} for precise definitions of the parameters. 

Now we would like to    clarify the term ``instantaneous" in the given context. One should consider several different time scales relevant for the problem. 
 First of all, there is a time scale $\Delta t_1$ which describes the  initial acceleration of emitted positrons with non-relativistic velocities up to the moment when it reaches the relativistic velocity $\simeq c$ along the direction of the ${\cal{E}}$-field. The corresponding time scale can be estimated as   $\Delta t_1\simeq (\frac{mc}{|e|{\cal{E}}} )\sim 0.01\mu s$. The next stage of acceleration is the motion of the positrons with relativistic speed along the electric field $\Delta t_2\simeq l_a/c\simeq 0.3 \mu s$.
 These two time scales   should be compared with time scale $\Delta t=\Delta r/c $  which determines the time spread of the arriving particles and recorded  as TASD events. The corresponding time scale can be estimated as $\Delta t \simeq \Delta s/c\simeq 3 \mu s$, see section \ref{confronts} for precise relations.  This  hierarchy of scales when $\Delta t\gg (\Delta t_1, \Delta t_2)$ implies that the corresponding times scales $\Delta t_1, \Delta t_2$ can be ignored in analysis of measured time spread for each given event within bursts, which justifies our treatment of the initial stage of the positron's motion as instantaneous.

 Next  important  parameters to consider  represent  the relevant time scales, which we are about to discuss. First, $\Delta t_{\rm burst}\lesssim 1$ ms represents  a maximum  time duration  for a single burst which is  treated in this work  as the cluster of  shower-like events from one and the same AQN. Second, each event within the burst does not normally lasts for more than  $\Delta t_{\rm event}\lesssim 10 {\mu s}$.  These time scales  can be represented in terms of the  corresponding distances travelled by AQN:     
   \be
\label{L1}
L_{\rm burst}\simeq v_{\rm AQN} \cdot  \Delta  t_{\rm burst} \simeq 250~ {\rm m}, ~~~ L_{\rm event}\lesssim 2.5 \rm ~m. ~~~
\ee
These time and length scales    play an important role in our comparison with  temporal and spatial features of pre-existing electric field under thunderstorms,
which is the topics of the next subsection \ref{cluster}.

The key element in these estimates is that the fluctuating electric field has sufficient correlation length $l_a$ and strength $\cal{E}$ which allow the positrons to accelerate to very large energies\footnote{It is interesting to note that according to \cite{DWYER2014147}  the positrons play a key role in development of the avalanche in RREA framework due to much longer mean free path  in comparison with electrons. The nature  of positrons in our framework  and in  \cite{DWYER2014147}   is  completely different, of course.}. Furthermore, this outbreak occurs on the time scale (\ref{parameters1}) determined by the properties of the electric field
while  the AQN itself remains at the same location  as $L_{\rm event}\ll l_a$ according to (\ref{L1}).

At the same  time  the scale $L_{\rm burst}\gg L_{\rm event}$ determines the number of   distinct  events which could potentially occur within the same burst during 1 ms.   
The number of possible  events is entirely determined by   the features of pre-existing electric field along the AQN's path. As we discussed above the field  strongly fluctuates temporally   and spatially with parameters (\ref{parameters1}). The event will  be recorded  by TASD if the   electric field points in the direction of the detector  within solid angle 
  $\Omega$  subtended at the instant location of the nugget when emission occurs. The solid angle $\Omega\sim 1$ is always  sufficiently large for distances $r\sim 10$ km. It remains true  even for relatively large zenith angles (skim events).

We now estimate the portion $\eta$    of the affected positrons at the instant when the AQN enters the region of  
strong electric field $\sim {\cal{E}}$.  The idea is to estimate  the ratio  of the positrons with binding energies smaller than  $\Delta E$ in comparison with total number of   positrons with binding energy exceeding $\Delta E$ (which could not be liberated by electric field). Assuming the Boltzmann distribution  when the  typical binding energy is order of $T$ the 
estimate for $\eta$ reads:
\be
\label{eta}
\eta\sim \frac{\int_{-\Delta E}^{0} d \epsilon \sqrt{|\epsilon|}\exp{[-\frac{\epsilon}{T}}]}{ \int^{-\Delta E}_{-T} d \epsilon \sqrt{|\epsilon|}\exp{[-\frac{\epsilon}{T}}]}\sim \left(\frac{\Delta E}{T}\right)^{\frac{3}{2}}\sim 0.1\left(\frac{10 \rm keV}{T}\right)^{\frac{3}{2}},~~~~~
\ee
where $\Delta E$ is proportional to electric field $\sim {\cal{E}}$ as given by (\ref{Delta_E}). 
One should emphasize that this estimate should be considered as  an order of magnitude at the best due to a number of uncertainties. First,   the excited weakly coupled positrons are not in the thermodynamical equilibrium and their actual distribution is not known. This is because the positrons get excited as a result of 
collisions with surrounding molecules and turbulent dynamics when AQN moves with large Mach number $M=v_{\rm AQN}/c_s$. Secondly, the parameter $\Delta E$ strongly depends on the dynamics of the 
electric field as discussed in subsection \ref{sect:thunder}. Both these components have large uncertainties. As a result, parameter $\eta$ carries a similar uncertainty,   
 which is obviously very hard to estimate as a clean theoretical description for both phenomena (turbulence and thunderstorm's electric field)  is not available at present time. 

   In next subsection we discuss the most puzzling observational feature of the bursts when at least three air shower triggers were recorded within 1 ms.
  We treat these (naively independent) events as a single cluster  generated by one and the same AQN traversing the Earth's atmosphere  with typical galactic dark matter   velocity $v_{\rm AQN}\sim 10^{-3}c$.

  \subsection{Mysterious bursts as the clustering events}\label{cluster}
  The main normalization factor  for our proposal is the maximal number of particles (positrons) $ N^{\rm max}_{\rm positrons}$ which can be emitted by AQN (and which can  potentially  reach TASD).
  The total number of positrons which have been accumulated by the AQN at the altitude around 10 km when the internal temperature reaches $T\simeq 10 $ keV is determined by formula (\ref{Q}). If a small portion $\eta$ of  the weakly coupled positrons will be suddenly liberated\footnote{\label{mu}The electrosphere contains   much more strongly coupled positrons characterized by very large chemical potential $\mu\simeq 20~ {\rm MeV}$. These strongly coupled  positrons cannot be liberated and they are ignored in the present discussions. Also, the rate of the annihilation processes with surrounding atmospheric material is extremely high  such that all liberated positrons will be quickly replaced as the temperature continues to rise with corresponding increase of  charge $Q$  according to (\ref{Q}).}   by the pre-existing thundercloud electric field  the maximal number of particles which can be detected by TASD localized at distance $r$ is estimated as  follows:
  \be
  \label{positrons}
  &&N^{\rm max}_{\rm positrons}\simeq \frac{(\eta \cdot Q)}{4\pi r^2} [507\cdot 3 {\rm m^2}]\cdot  \la e^{-\frac{r}{\lambda}}\ra\\ &\simeq & 10^3\cdot \left(\frac{\eta(T)}{0.1}\right)\cdot \left(\frac{T}{10~\rm  keV}\right)^{5/4} \cdot \left(\frac{10~\rm  km}{r}\right)^2\cdot\left(\frac{ \la e^{-\frac{r}{\lambda}}\ra}{0.1}\right), \nonumber
  \ee
  where we substitute the total area of the detector as 507 SDs with area $3~ \rm  m^2$ each. The  detected number of particles for a  shower-like event within the same cluster (burst) is   of order $10^3$ or less   according to \cite{Abbasi:2017rvx,Okuda_2019}, which is consistent with estimate (\ref{positrons}).    
  
 After the positrons are liberated from the nugget, they will be accelerated to  the energies of order 10 MeV by pre-existing electric field with typical scale of order $\rm kV/cm$ according to the results of previous subsection \ref{sect:thunder}.   In our estimate (\ref{positrons}) we assumed that the mean free path $\lambda$ for positrons with few MeV energy  is order of kilometre at the sea level and several  kilometres at higher  altitudes which   gives a  suppression factor $\la \exp(-r/\lambda)\ra\sim 0.1$ for  particles with  energies in few MeV range. No much  suppression  occurs for  higher energy positrons with energies 10 MeV or higher, which is often the case as argued in subsection \ref{sect:AQN-thunder}. 
  
  In  (\ref{positrons}) we also assumed that $\eta\simeq 0.1$ which describes a  portion of the liberated positrons by preexisting electric   field in thunderclouds.   The corresponding suppression factor is very hard to estimate on the quantitative level as it is  determined   
  by non-equilibrium  dynamics as explained in the text below    (\ref{eta}). The order of magnitude estimate given in  subsection \ref{sect:AQN-thunder}
  is consistent with $\eta\simeq 0.1$. This portion of positrons will be replaced (fast refill) very quickly due to the very fast   processes, see footnote \ref{mu}.

\exclude{Second, the positrons  (\ref{positrons})  saturating   the shower-like events within the same  burst must be emitted from this small portion of the AQN trajectory $L_{\rm burst}$. 
This length scale is obviously much smaller than 10 km which is a typical distance between TASD and  AQN trajectory. Therefore, our estimation (\ref{positrons}) that the burst emission occurs from a point-like object is justified. This scale also shows that the relevant portion of the AQN trajectory
emitting the positrons in the TASD direction always stays in the same thundercloud which normally is much larger in size. 
}

We conclude this section with the following generic remarks. First, all estimates presented in this section are based on specific features of the electric field (\ref{parameters1}) which are known to be present in thunderclouds. The same fields are  known to be responsible for the lightening processes as well, which of course much more frequent and numerous events. However, the mechanism described above is not  literally associated with lightening flashes. In particular, Fig 9 in \cite{Abbasi:2017rvx} explicitly shows that 
some SD events occur earlier than lightning. Furthermore, some events are not related to the lightnings at all.   These observations  unambiguously imply  that  the ``mysterious bursts"    are associated  with very initial processes (such as generation of strong electric field under thunderstorms), but not with  lightning flashes themselves. 

Our second remark goes as follows. The main claim of this work    are  based on  analysis of the  unique features listed as $\bf 1-\bf 7$  of the events,  and which are hard to explain within conventional CR assumption. Some of the observables 
as discussed in next Sect. \ref{confronts}   have pure geometrical and kinematical nature and represent essentially model independent predictions within AQN proposal, in contrast with the estimation (\ref{positrons})   which suffers  large numerical uncertainties as explained above.

\section{``Mysterious bursts" proposal \ref{proposal} confronts the observations \cite{Abbasi:2017rvx,Okuda_2019}}\label{confronts}
The goal of this section is to confront the theoretical ideas  of the proposal (``Mysterious bursts as the AQN annihilation events"  formulated  in section  \ref{proposal}) with observations  \cite{Abbasi:2017rvx,Okuda_2019}. We explain how the unusual features from  items {\bf 1-7} listed in section \ref{sec:introduction} can be naturally understood within the AQN framework.

We start our discussions with item {\bf 1} from the list. In the AQN framework all positrons are emitted from 
the typical  altitude   for thunderstorm which is around 10 km. This altitude  is obviously much lower than 30 km   when CR air shower normally starts. Furthermore, 
the events appear to be much more ``curved" , see Figs. 3 and  4 in  \cite{Abbasi:2017rvx}.
In the AQN framework this ``curved" feature can be easily understood by noticing that the essential parameter in this proposal is the initial spread of the particles 
determined by angle $ \Delta \alpha\simeq \left( {v_{\perp}}/{c}\right)$ as given  by (\ref{spread}). 
  Indeed, the parameter $\Delta t$ can be estimated as follows:  
  \be
\label{spread1}
 \Delta t \simeq  \frac{\Delta r}{c}\simeq 3 \mu s \cdot (\tan\alpha)\cdot 
   \left(\frac{r}{10 \rm ~km}\right) \left(\frac{\Delta \alpha}{0.1}\right) ~ {\rm where}
 ~\Delta r \simeq  r \tan\alpha \Delta \alpha\simeq 1~{\rm km} \cdot (\tan\alpha)\cdot 
   \left(\frac{r}{10 \rm ~km}\right) \left(\frac{\Delta \alpha}{0.1}\right),  ~~~
      \ee
see Fig. \ref{geometry} with definitions of all the parameters. 
One can infer from (\ref{spread1}) that  time spread $(2\Delta t)$ indeed could be relatively large, up to $  (7-8) \mu s$, 
as observed,  see Fig. 3 and Fig. 4 in \cite{Abbasi:2017rvx}. 
It should be  contrasted with conventional CR air shower when this timing spread  is always below $2 \mu s$, see Fig. 5 in \cite{Abbasi:2017rvx}. We use $(2\Delta t)$ in our estimates with extra factor 2 as  the angle $\Delta \alpha=(v_{\perp}/c)$    determining  the particle distribution,    could assume the  positive or  negative value, depending on sign of $v_{\perp}$ with respect to instant direction of the electric field $\vec{{\cal{E}}}$ as shown on  Fig. \ref{geometry}. 
 
Another distinct characteristic of the the AQN framework  is much stronger localization of the particles around the axis when it is typically below 2km,  see Fig. 3 and Fig. 4 in \cite{Abbasi:2017rvx}, in contrast with conventional CR air shower when it normally assumes much larger values around 3.5 km, Fig. 5 in \cite{Abbasi:2017rvx}. This feature also finds its  natural explanation within the AQN framework. Indeed, the spread of the particles on the plane determined by $\Delta s$, while the timing is characterized by $\Delta t \simeq {\Delta r}/{c}$ estimated above (\ref{spread1}). These parameters are linearly proportional to each other and reside in proper range of parameters consistent with observations. Indeed,    
\be
\label{spread2}
\Delta s \simeq  r \left(\frac{\Delta \alpha}{\cos \alpha}\right) \simeq \frac{1~{\rm km}}{\cos\alpha} \left(\frac{r}{10 \rm ~km}\right) \left(\frac{\Delta \alpha}{0.1}\right), ~~~ \Delta r \simeq \Delta s \sin\alpha , ~~~\Delta \alpha\simeq \left(\frac{v_{\perp}}{c}\right)\in (0- 0.1)
\ee
such that $2\Delta t$ may vary between $(0.5-8) \mu s$ when $2\Delta s$ changes between $(0.5-2)$ km with approximately linear slope determined by electric field direction $\sin \alpha$. This behaviour is    consistent with observed events presented on Fig. 3 and Fig. 4 in \cite{Abbasi:2017rvx}. 

Similar  arguments also explain why the observed events do not have sharp edges in waveforms  (see Fig. 6 in  \cite{Abbasi:2017rvx}) as  listed in item {\bf 2}. The point is that conventional CR  air showers typically  have a single ultra-relativistic particle which generates a very sharp edge in waveforms. It should be contrasted with large number of positrons in our proposal explaining TASD bursts. These positrons are  characterized by initial 
transverse velocity distribution at the moment of exit, and cannot be thought as a single ultra-relativistic particle.   Therefore, these multiple non-coherent positrons cannot generate   sharp edges in waveforms which are normally observed   in case of conventional CR events. Instead, these positrons produce the non-sharp edges in waveforms.

The AQN traverses   the distance $L_{\rm burst} $   according to (\ref{L1}) during $\Delta t= 1 $ ms representing the {\it cluster}  of events in the AQN framework.   This distance  never exceeds 1 km. Furthermore, the spatial spread for each individual   event within the same cluster also within the same range $\sim$ 1 km according to (\ref{spread}). These estimates are perfectly consistent with item {\bf 3} which is extremely hard to explain in terms of conventional CR air showers. 

The item {\bf 4} finds a very natural explanation within AQN proposal. Indeed, the intensity of the events within the burst is determined by (\ref{positrons}) with the number of particles corresponding to $(10^{18}-10^{19})$ eV energy range if analyzed in terms of     conventional CR air shower event. 
However, in the AQN framework the number of particles is determined by the different parameters, such as internal temperature\footnote{\label{parameters}The corresponding AQN's properties such as baryon charge $\la B\ra$, the temperature $T$  and the  $Q$ have  been previously computed for completely different purposes in different context. By no means we fitted these parameters  to accommodate the observations.}
 of the nugget $T$, while the burst is considered to be the cluster of events originated from one and the same AQN  traversing very short      distance $L_{\rm burst} $ during 1 ms. 

According to  item {\bf 5}   all observed bursts occurred under thunderstorm.  It is very hard to interpret this feature in terms of conventional CR as the conventional air showers cannot be drastically modified as a result of the thunderstorm. However, it is  perfectly consistent with AQN framework 
because the thunderstorm with its pre-existing electric field (\ref{E}) plays a crucial role in our mechanism as the electric field instantaneously liberates the positrons and also accelerates them up to 10 MeV energies.     

 It has been observed that all bursts occur at the same time of lightning or earlier than lightning, according to item {\bf 6}, see Fig. 9 in  \cite{Abbasi:2017rvx}. 
 Some bursts are not related to lightning at all. This observation unambiguously implies that  the bursts are associated  with processes  which were present (such as fluctuating electric field) before lightning flashes may (or may not) occur. 
 The AQN mechanism obviously satisfies this requirement   as large number of particles (\ref{positrons}) have been prepared long before the lightening flashes, and the electric field commonly present in thunderstorm plays role of a trigger liberating   the large number of the particles when the AQN enters the background  electric field under thunderstorm. 
 
One can naively think that one should expect much more bursts which are not correlated with lightnings. 
  The basic reason why such strong correlation occurs within AQN model is that both phenomena (the lightnings and bursts) are not independent events as they are both triggered by the same strong electric field which is present in the system. When this field locally assumes a sufficiently large values with ${\cal{E}}\gtrsim {\cal{E}}_c$ exceeding the critical values, the lightning starts as a result RB \cite{Gurevich_2001} or RREA \cite{DWYER2014147} processes. A similar  sufficiently  large field ${\cal{E}}\sim {\cal{E}}_c$ (which may or may not generate a lightning) plays a crucial role liberating  the positrons in the same local area at the same instant. A precise estimations are very hard to make as the dynamics of the small scale  strong  fluctuating electric field 
 [determined by   (\ref{E}) and (\ref{l_a})]  which  is initiating the lightning is not well understood\footnote{It should be contrasted with large scale electric field between different clouds and cloud and ground with typical scale measured in km with typical time scales measured in minutes, in contrast with parameters (\ref{E}) and (\ref{l_a}) characterizing small scale local instantaneous field. Large scale structure which is irrelevant for our purposes can be properly described by e.g. tripole model  \cite{DWYER2014147}.}.  
 
 It is also very possible that AQNs play the role of the triggers as mentioned in footnote \ref{trigger} replacing the  conventional CR which normally serve as  the triggers by starting the lightnings according to  \cite{Gurevich_2001}.    It is important  for our purposes that the values of the field which is capable to liberate positrons from AQNs and initiate 
the lightning assume the same order of magnitude on the level (\ref{E}).

 The item {\bf 7} also finds its    natural explanation within AQN proposal. The frequency  of appearance of these ``mysterious bursts"  in our framework  is determined by (\ref{N1})
 which is marginally  consistent with the observed 10 events  \cite{Abbasi:2017rvx,Okuda_2019}.  
 Once again, the parameters which control  the frequency of these events are mostly determined by the dark matter flux (\ref{eq:D Nflux 3})  expressed in terms of  the AQN size distribution. These parameters have been fixed long ago for completely different purposes in different context. We did not  attempt to  fit these parameters  to accommodate the observations \cite{Abbasi:2017rvx,Okuda_2019}. 
 
 Furthermore, the injection energy $\Delta E$ as given by (\ref{Delta_E}) is expressed in terms of thunderstorm parameter $\cal{E}$ and also in terms of $T, Q$ which were computed long ago irrespectively to thunderstorm physics. Nevertheless, the obtained value for $\Delta E$ in keV energy range is precisely what is required to liberate the positrons and consequently accelerate them, see also footnote \ref{parameters}. 
 
 To summarize this section: there are two types of observables  being  discussed in this work.
 First type of  observables  such as   frequency  of appearance (\ref{N1}) and    intensity of the events within the bursts (\ref{positrons})
 essentially describe absolute normalizations, and 
 are highly sensitive to the parameters of the system such as  thunderstorm electric field $\cal{E}$ or $\eta$ parameter which are hard to compute from the first principles as explained after  (\ref{eta}). 
 
 There is  another  type of  observables which are essentially insensitive to these parameters and do not suffer from corresponding uncertainties. 
 These observables have pure geometrical and kinematical nature and represent essentially model independent prediction within AQN proposal.
  In particular,  the   time spread scale $\Delta t$ characterizing each given event (\ref{spread1}).  Another parameter is the spatial spread $\Delta s$ 
  over SD area as given in (\ref{spread2}). Both these parameters are not very sensitive to any uncertainties mentioned above as they expressed in terms of pure kinematics determined by a small transverse kick $v_{\perp}$ carried by the positrons at the point of exit.  The relation between these two observables is also insensitive to  any uncertainties mentioned above, but is determined exclusively  by the instant direction of the local electric field $\cal{E}$ where positrons have been liberated along    the AQN path, see Fig. \ref{geometry}.  These features are perfectly consistent with item {\bf 1}   from the list 
  as formulated  by the authors \cite{Abbasi:2017rvx,Okuda_2019}  in terms 
  of  higher  curvature of the burst events in comparison with conventional  CR air showers. 
 
 \exclude{
 We consider the multiple  ``numerical coincidences" listed above as  strong supporting arguments  for this proposal
 as all recorded  features such as   frequency  of appearance (\ref{N1}), the observed  intensity of the events within the bursts (\ref{positrons}) as well as required value $\Delta E$ being  in keV energy range  are all expressed in terms of parameters covering very different fields  of  physics which span enormous  range of scales.   These scales include but not limited: the DM density $\rho_{\rm DM}$,  thunderstorm electric field $\cal{E}_c$,  microscopical parameter $T$, to name just few. 
 All these parameters have many uncertainties as discussed in the text.

  as well as the distance surface distribution characterized by parameter $\Delta s$ as given in (\ref{spread2}) are also estimated in terms of the same set of parameters as $\Delta \alpha$ is expressed in terms of $v_{\perp}$ which itself is determined by internal temperature $T$.
 It is very nontrivial self-consistency check  that  these parameters   [from very distinct fields of physics, being fixed  by fundamentally different observations]    nicely ``conspire"   to produce very reasonable numerical estimates which are consistent with puzzling  features  {\bf 1-7} as recorded by \cite{Abbasi:2017rvx,Okuda_2019}. 
}

\section{Conclusion}\label{conclusion}
Our basic results can be summarized as follows.
We argued that the mysterious  bursts (with highly unusual features as listed by items {\bf 1-7} in Introduction)  of shower-like events observed by TASD  \cite{Abbasi:2017rvx,Okuda_2019}  are naturally interpreted as the cluster events generated by the AQNs propagating in thunderstorm environment. We presented our arguments   in section \ref{confronts} where we explained how the puzzling features  {\bf 1-7}, item by item,  naturally emerge in the AQN framework. There is no need to repeat these arguments again here in Conclusion, and  we refer to the two last paragraphs  of the previous section \ref{confronts}   for  the  summary. Instead, we would like to describe three   specific tests of this framework which can confirm, substantiate or refute our proposal.

$\bullet$ First of all, the time scale $\Delta t_{\rm burst}=1$ ms as the definition for the burst is  obviously  an ad hoc parameter. We suggest to reanalyze the existing  data to increase this parameter by factor (2-4) or event up to $\Delta t_{\rm burst}=10$ ms. 
We would like  to see if more events will be recorded within the  ``prolongated  burst", and more new bursts will emerge which  previously were not qualified as the bursts
(because they had less than three consecutive  events).  

Our proposal predicts that the answer on both questions should be positive: there should be more events in previously recorded bursts, and  it should be  more    new bursts being recorded if $\Delta t_{\rm burst} $ to be increased. In fact, it has been mentioned in  \cite{Abbasi:2017rvx}
that there are several two-event bursts and many single events with   features similar to the events within bursts.  However they  were not qualified as the bursts. It would be interesting to see if these previously ``unqualified" events become ``qualified" events for the  ``prolongated  burst".
  The  existing cluster may become larger by accommodating  new  events within the ``prolongated  bursts" and 
  new clusters  may also emerge.

One should emphasize that this is a highly nontrivial prediction based on many specific features of our mechanism where the bursts are the cluster events generated by one and the same AQN traversing the thunderclouds in the area with typical DM velocity $\sim 10^{-3}c$. Therefore, its entire trajectory  for ``prolongated  bursts"  remains in the same area within   $L_{\rm burst}\lesssim$ 2.5 km scale even for $\Delta t_{\rm burst}\simeq 10$ ms.

\exclude{ 
The interpretation in terms of conventional CR would remain very problematic even for larger $\Delta t_{\rm burst}$ as the chance of coincidence would  remain to be very tiny.  In fact the corresponding probability  (estimated in  \cite{Abbasi:2017rvx}) could become even smaller as the number of qualified bursts likely to increase for ``prolongated  bursts" with  $\Delta t_{\rm burst} > 1 $ ms.
}

One should also note that the expansions plans  \cite{Bergman:2020izr}  to increase the area up to 3000 $\rm km^2$  to include 500 new SD counters would be a highly beneficial element for the present proposal as it should  increase the frequency  of appearance  of qualified bursts to be observed with new facilities. 
It is also very important that kinematical, essentially model independent,  relations (\ref{spread1}) and  (\ref{spread2}) can be quantitatively tested  with more statistics.  
Precisely the puzzling ``curved" features of the burst events can be understood in terms of (\ref{spread1}) and  (\ref{spread2}).

 $\bullet$ The second possible test  we advocate  is to study the  unique correlations  with acoustic and seismic  signals which is always accompany the propagation  of the AQNs in atmosphere. Infrasound, acoustic and seismic signals were studied 
  in  \cite{Budker:2020mqk}. 
  \exclude{The main topic of  that paper  was   the analysis of the infrasound and seismic acoustic waves  which were recorded by Elginfield Infrasound Array (ELFO) and seismic stations
 near London, Ontario, Canada
   on July 31st 2008. It   was considered as a very mysterious event as it was very different from conventional meteor like events as the synchronized  all sky cameras did not observe any activities at that time. 
 It has been shown in \cite{Budker:2020mqk} that the intensity and the frequency of   the infrasound and seismic   waves are consistent with observed records if this event is interpreted in terms of the AQN framework. However, the recorded intensity of the infrasound signal corresponds to vey large size  AQN with $B\simeq 10^{27}$, which are very rare events (once in 10 years). 
   }  
    It has been  proposed a detection strategy to search 
 for infrasound and seismic  signals    by using Distributed Acoustic Sensing  (DAS) instruments. Furthermore, it has been shown that 
     using an amplifier chain one can extend the range of DAS unit to 82 km, while maintaining high signal quality.   The new element we are advocating in the present work is as follows. One can use the existing technology with  DAS and install it  in the same   location where   TASD stands. It would allow to study the correlations  between the bursts (interpreted as the cluster events within AQN framework) and small seismic event signals   which would be recorded by DAS.  An important point here  is that DAS  can in principle  detect not only the intensity and the frequency of the sound wave, but also the direction of the source. This direction can be cross correlated with TASD which is also capable to reconstruct the source of the burst event. 
 
 $\bullet$ The third test we advocate is to study the unique correlations of the TASD events with the radio signals which will   always  accompany the acceleration of the liberated positrons.   It is absolutely inevitable feature of the system and unique prediction of the AQN framework. One can show \cite{arz-radio} that the frequency of radio waves is mostly determined by the avalanche length scale $l_a$ such that the frequency of the radio  emission is of order 
  $\nu\in (3-300)$ MHz. A very short radio pulse must arrive to SD area within few $\mu s$ from TASD events (depending on exact position of the radio  antenna) 
because   the positrons and radio waves propagate with the same speed $c$ and they both emitted  at the same instant from the same location. 
This synchronization must hold irrespectively whether TASD events synchronized, related  or not  related with the lightning. 
 
  Our  present proposal suggests  that the bursts with very unusual features as recorded by TASD \cite{Abbasi:2017rvx,Okuda_2019} may be  in fact  the AQN annihilation events under thunderstorm.  If  this interpretation is confirmed by future studies
 it would be the    {\it first~direct }  (non-gravitation) evidence which reveals  the nature of the DM.

   \section*{Acknowledgements}

This research was supported in part by the Natural Sciences and Engineering
Research Council of Canada. 

 \appendix
 \section{The AQN internal temperature under thunderstorm }\label{sect:temperature} 
 The goal of this Appendix is to overview the basic  characteristic of the AQNs, the   internal temperature $T$ which enters several formulae in  Section \ref{TQ} in the main text. The corresponding computations have been carried out in \cite{Forbes:2008uf}
 in application to the galactic environment with a typical density of surrounding visible baryons of order $n_{\rm galaxy}\sim 300 ~{\rm cm^{-3}}$ in the galactic center. We review  these computations with few additional elements which must be implemented for Earth's atmosphere when 
 typical density of surrounding baryons is much higher $n_{\rm air}\simeq 30\cdot N_m\simeq 10^{21} ~{\rm cm^{-3}}$, where 
 $N_m\simeq 2.7\cdot 10^{19}  ~{\rm cm^{-3}}$ is the molecular density  in atmosphere when each molecule contains approximately 30 baryons. 

The total surface
emissivity from electrosphere has been computed in \cite{Forbes:2008uf} and it is given by 
\begin{equation}
  \label{eq:P_t}
  F_{\text{tot}} \simeq 
  \frac{16}{3}
  \frac{T^4\alpha^{5/2}}{\pi}\sqrt[4]{\frac{T}{m}}.\\
\end{equation}
 
A typical internal temperature of  the nuggets can be estimated from the condition that
 the radiative output of equation (\ref{eq:P_t}) must balanced the flux of energy onto the 
nugget  due to the annihilation events. In this case we may write, 
\be
\label{eq:rad_balance}
F&\simeq &\left[(4\pi R^2) \frac{16}{3}
  \frac{T^4\alpha^{5/2}}{\pi}\sqrt[4]{\frac{T}{m}}\right] +\Delta F_{\rm other}\\
&\simeq& \kappa\cdot  (\pi R^2) \cdot (2~ {\rm GeV})\cdot n_{\rm air} \cdot v_{\rm AQN}, \nonumber 
\ee 
where the left hand side accounts for the total energy radiation from the nuggets' surface per unit time as given by (\ref{eq:P_t}) plus other processes  denoted as $\Delta F_{\rm other}$ to be discussed below. 
 The right hand side  accounts for the rate of annihilation events when each successful annihilation event of a single baryon charge produces $\sim 2m_pc^2\simeq 2~{\rm GeV}$ energy. In  (\ref{eq:rad_balance}) we assume that  the nugget is characterized by the geometrical cross section $\pi R^2$ when it propagates 
in environment with local density $n_{\rm air}$ with velocity $v_{\rm AQN}\sim 10^{-3}c$.

The factor $\kappa$ is introduced to account for the fact that not all matter striking the nugget will 
annihilate and not all of the energy released by an annihilation will be thermalized in the nuggets. In particular,  some portion of the energy will be released in form of the axions, neutrinos and  liberated positrons by the mechanism  discussed in the main text of this work. 
This portion is  represented by $\Delta F_{\rm other}$. The parameter $\kappa$ was estimated for the galactic environment in \cite{Forbes:2008uf}. This parameter obviously must be different for the earth's atmosphere. However, for the order of magnitude estimates we ignore this difference. 
 
As such $\kappa$ encodes a large number of complex processes including the probability that 
 not all atoms and molecules   are capable to  penetrate into the 
color superconducting phase of the nugget to get annihilated. Furthermore, some positrons can be liberated due to the pre-existing electric field in thunderclouds as discussed in the main text of this work. Furthermore, there is another complication as the AQN moves with supersonic velocity. This generates the shock waves and turbulence in earth's atmosphere, which makes the computations even more complicated. 

 In a neutral dilute environment considered  previously \cite{Forbes:2008uf}  the value of $\kappa$ cannot exceed $\kappa \lesssim 1$ which would 
correspond to the total annihilation of all impacting matter into to thermal photons. The high probability 
of reflection at the sharp quark matter surface lowers the value of $\kappa$. The propagation of an ionized (negatively charged) nugget in a  highly ionized plasma (such as solar corona)   will 
 increase  the effective cross section. As a consequence,   the value of  $\kappa$ could be very large as discussed in \cite{Raza:2018gpb} in application to the solar corona heating problem.
 
 The propagation of the AQNs under thunderstorm,  which is the topic of the present studies, is an intermediate case between these two previous studies. 
 We shall argue below that ionization effects can be ignored in this case. Furthermore, extra term $\Delta F_{\rm other}$ related to the positron's liberation under the thunderstorm (which is   a new effect not considered previously)   also does not modify our  estimates.   Therefore,  one can estimate a typical internal nugget's temperature in the Earth atmosphere at altitude $\sim 10$ km as follows:
 \be
 \label{T}
 T\simeq 15 ~{\rm keV} \cdot \left(\frac{N_{\rm m}}{10^{19} ~{\rm cm^{-3}}}\right)^{\frac{4}{17}}\left(\frac{ \kappa}{0.1}\right)^{\frac{4}{17}},
 \ee 
which represents a typical internal temperature of the AQNs relevant for the present work. All the uncertainties  related to $\kappa$ mentioned above do not modify our qualitative discussions in this work. 
 
 In this rest of the Appendix we would like to argue that the ionization features of the nuggets and the environment do not modify our estimates.
 First, we estimate the kinetic energy of the molecules in the AQN frame:
 \be
 E_{\rm kinetic}\rm \simeq \frac{(30 ~GeV) \cdot v^2_{\rm AQN}}{2}\simeq 15 ~{\rm keV}, 
 \ee
 which is numerically the same order of magnitude as the internal temperature of the nuggets (\ref{T}). Therefore, the typical distance where the positive ions  (which is always present under the thunderstorm) can modify the annihilation rate  assumes the value estimated in (\ref{capture1}) and coined as the $R_{\rm cap}$.
 
 The density of ions $N_i$ under normal conditions is estimated as $N_i\simeq 10^3 {\rm cm^{-3}}$ while under RB conditions it could be as high as  $N_i\simeq (5-8) \cdot 10^4 {\rm cm^{-3}}$ according to  \cite{Gurevich_2001}. Therefore, 
 the extra term due to the ionization features of the environment can be represented as follows
 \be
 \rm \left(N_m R^2+ N_i R_{\rm cap}^2\right)=N_m R^2\left[1+\frac{N_i}{N_m}\left(\frac{R_{cap}}{R}\right)^2\right].~~
 \ee
 One can check  that the correction proportional to $N_i$ numerically is at least 4 orders of magnitude smaller than the main term, and therefore, can be ignored, 
 which a posteriori justifies our approximation.  
 
 Now we would like to argue that   another assumption we made when we ignored $\Delta F_{\rm other}$ related to the liberated positrons (due tot the presence of electric field during the    thunderstorms) is also justified. Indeed, the number of direct collisions of the molecules with AQN per unit time can be estimated as follows:
\be
\label{collisions}
\frac{dN_{\rm collisions}}{dt}&\simeq& (\pi R^2) \cdot   N_{\rm m} \cdot v_{\rm AQN}\\
&\simeq& 0.3\cdot 10^{18} \left(\frac{N_{\rm m}}{10^{19} ~{\rm cm^{-3}}}\right)\rm s^{-1}.\nonumber 
\ee   

The dominant portion  of these collisions are the elastic scattering processes rather than successful annihilation events  suppressed by parameters $\kappa$ as discussed above. Some of these elastic collisions lead to the energy transfer   to the positrons from electro-sphere, which also relatively rare events. Some positrons, may even leave the system as a result of the collisions with rate  (\ref{collisions}) under influence of the strong electric field as discussed in this work.  However, even if one assumes that every single collision event (\ref{collisions}) liberates a positron as a result of electric field lasting a $\mu s$ one would get 0.5 MeV energy lost per a single collision, while the annihilation energy gain is 3 orders of magnitude higher as it is proportional $\kappa$ GeV per collision.  Therefore, the term $\Delta F_{\rm other}$ which enters the equation for the energetic balance (\ref{eq:rad_balance}) and related to the liberated positrons indeed can be ignored. 
 This justifies our approximation  a posteriori.

\bibliography{Bursts-revised}

\end{document}